\documentclass[12pt]{article}
\begin{document}
\begin{center}
\LARGE{On Isosceles Triangles and Related Problems in a Convex Polygon}
\end{center}
\large{
\begin{center}
Amol Aggarwal \\ 
Saratoga High School \\ 
Saratoga, California \\
June 19, 2010
\end{center} }

\normalsize{
\begin{abstract}
Given any convex $n$-gon, in this article, we: (i) prove that its vertices can form at most $n^2/2 + \Theta(n\log n)$ isosceles trianges with two sides of unit length and show that this bound is optimal in the first order, (ii) conjecture that its vertices can form at most $3n^2/4 + o(n^2)$ isosceles triangles and prove this conjecture for a special group of convex $n$-gons, (iii) prove that its vertices can form at most $\lfloor n/k \rfloor$ regular $k$-gons for any integer $k\ge 4$ and that this bound is optimal, and (iv) provide a short proof that the sum of all the distances between its vertices is at least $(n-1)/2$ and at most $\lfloor n/2 \rfloor \lceil n/2 \rceil(1/2)$ as long as the convex $n$-gon has unit perimeter.
\end{abstract}

\section{ Introduction} 
\noindent In 1959, Erd\"{o}s and Moser asked the following question in [11]: What is the maximum number of unit distances that can be formed by vertices of a convex $n$-gon?  They conjectured that this bound should be linear, and in [9], Edelsbrunner and Hajnal provided a lower bound of $2n-7$.  On the other hand, F\"{u}redi provided an upper bound of $2 \pi n \log_2 n - \pi n$ in [12], and recently in [6], Brass and Pach gave an upper bound of $9.65 n \log_2 n$ using induction and geometric constraints different from those provided by F\"{u}redi. These bounds were later improved to $n\log_2 n + 4n$ in [2].  \\ 

\noindent In [3], Altman proved that the number of distinct distances among all of the vertices of any convex $n$-gon is at least $\lfloor n/2 \rfloor$, a bound that is achieved by a regular polygon. Moreover, in [4], Altman proved several useful properties about the lengths of the diagonals of convex $n$-gons. Dumitrescu showed in [7] that at most $(11n^2 - 18n)/12$ isosceles triangles can be created by the vertices of a convex $n$-gon and uses this upper bound to show that there are at least $\lceil (13n-6)/36 \rceil$ distinct distances from some vertex, thereby making progress on Erd\"{o}s's conjecture in [10] that there is a vertex in a convex $n$-gon that is at distinct distances from at least $\lfloor n/2 \rfloor$ other vertices. In [17], Pach and Tardos showed that the number of isosceles triangles formed by a set of $n$ vertices in the plane is at most $O(n^{2.136})$.  In [1], \'{A}brego and Fern\'{a}ndez-Merchant showed that there are at most $n-2$ equilateral triangles that can be created by the vertices of any convex $n$-gon. Furthermore, in [16], Pach and Pinchasi showed that the number of unit distance equilateral triangles is at most $\lfloor 2(n-1)/3 \rfloor$, and they exhibit a convex $n$-gon for which this bound is achieved. \\ 	

\noindent Before we discuss the results of this paper, we define a few terms with regard to a convex polygon. Call an edge of a {\itshape unit edge} if the length of the edge is one and call a triangle a {\itshape unit isosceles triangle} if it has at least two unit edges. We call vertex $v$ a {\itshape centroid} if there exist three vertices, $v_1, v_2,$ and $v_3$ such that $d(v, v_1) = d(v, v_2) = d(v, v_3)$, where $d(u, v)$ is the Euclidean distance between two points $u$ and $v$ in the plane. The circle with center $v$ and radius $d(v, v_1)$ is one of $v$'s {\itshape centroid-circles}. Note that $v$ can have multiple centroid-circles. We say that two centroid-circles {\itshape intersect} if they share a vertex of the polygon, and call a centroid-circle {\itshape intersecting} if it intersects at least one other centroid-circle. In this article, we prove the following results.

\paragraph{Theorem 1:} There are at most $n^2/2 + \Theta(n\log n)$ unit isosceles triangles formed by vertices of any convex $n$-gon. \\ 

\noindent This bound is sharp in the first term because we exhibit a convex $n$-gon that forms $(n^2-3n+2)/2 + \lfloor (n-1)/3 \rfloor$ unit isosceles triangles.  

\paragraph{Theorem 2:} Suppose that $\mathcal{P}$ is a $n$-gon that has no centroid-circles that intersect. Then, there are at most $3(n+1)^2/4$ isosceles triangles formed by vertices of $\mathcal{P}$.  

\paragraph{Theorem 3:} Suppose that $\mathcal{P}$ is a convex $n$-gon that has $k$ intersecting centroid-circles with $k = o(n^{2/3})$. Then, there are at most $3n^2/4 + o(n^2)$ isosceles triangles formed by vertices of $\mathcal{P}$.  \\

\noindent In Section 3, we show that the there exists a convex $n$-gon that creates $(3n^2 - 11n + 8 + 2\lfloor n/2 \rfloor)/4$ isosceles triangles, meaning that these bounds are sharp in the first order.

\paragraph{Theorem 4:} Let $n$ and $k$ be integers greater than $3$.  The maximum number of regular $k$-gons that can be found in a convex $n$-gon is $\lfloor n/k \rfloor$ and this bound is sharp. 

\paragraph{Theorem 5:} For any convex $n$-gon with unit perimeter, the sum $S_n$ of distances between its vertices satisfies $(n-1)/2 \le S_n \le (1/2) \lceil n/2  \rceil \lfloor n/2 \rfloor$. \\

\noindent In Section 5, we show that the results of Altman in [4] can be easily used to prove a conjecture given in [5] by Audet, Hansen, and Messine regarding the sum of distances between the verticies of a convex $n$-gon with unit perimeter. This result has also been proven by Larcher and Pillichshammer in [13], and Dumitrescu later extends their proof to work for concave polygons in [8].

\section{Number of Unit Isosceles Triangles}

\paragraph{Proposition 1:} There exists a polygon that forms $\displaystyle\frac{n^2 - 3n + 2}{2} + \left\lceil \displaystyle\frac{n-1}{3} \right\rceil$ unit isosceles triangles.

\noindent \paragraph{Proof:} Consider vertices $v, v_1, v_2, v_3, \cdots v_{n-1}$ such that $v_1v_2\cdots v_n v$ is convex, $d(v, v_i) = 1$ for all $1\le i \le n -1$, and $d(v_i, v_{i+k}) = d(v_{i+k}, v_{i+2k}) = 1$, where $k = \lfloor n/3 \rfloor$ and $1\le i\le \lfloor (n-1)/3 \rfloor$. Then, $\triangle vv_iv_j$ is isosceles for any $1\le i < j \le n - 1$. Moreover, triangle $v_iv_{i+k}v_{i+2k}$ is isosceles for any $1\le i\le \lfloor (n-1)/3 \rfloor$. Thus, we have a total of $(n^2-3n+2)/2 + \lfloor (n-1)/3 \rfloor$ isosceles triangles. 
  \rule[0mm]{2.6mm}{2.6mm}% 

\paragraph{Theorem 1:} The number of unit isosceles triangles that can be formed by vertices of a convex $n$-gon is at most $n^2/2 + 4n\log n + 20n + 8$ for sufficiently large $n$.  
\paragraph{Proof:}  The idea of the proof is based on Dumitrescu's paper [8] and Moser's paper [14].  Let the convex $n$-gon be $\mathcal{P}$.  Consider the smallest circle that covers all vertices of $\mathcal{P}$.  At least two vertices of the polygon lie on this circle.  We examine two cases: one in which there are precisely two vertices on this circle and one in which there are at least three vertices on this circle.  \\

\noindent {\bfseries Case 1:} Only two vertices of $\mathcal {P}$ lie on this circle.  Then, these two vertices must form the diameter of the circle.  Let the polygon be $v_1v_2v_3\cdots v_n$ with $v_1v_k$ as the diameter of the circle.  Let the vertices $v_1, v_2, v_3, \cdots , v_k$ form set $S$ and let $v_k, v_{k+1}, v_{k+2}, \cdots , v_n, v_1$ form set $S'$.  Let $|S| = a$ and $|S'| = b$.  Then, $n + 2 \ge a + b$.  Consider any vertex $v_j$ in $S$.  For any $i, j$ so that $1 < i \le j \le k$, $\angle v_jv_iv_{i-1} \ge \angle v_1v_iv_k \ge \pi/2$, so $d(v_{i-1}, v_j) > d(v_{j}, v_i)$, implying that the distances from $v_j$ to the vertices between $v_1$ and $v_j$ in $S$ are all distinct.  By similar logic, the distances between $v_j$ and vertices between $v_{j+1}$ and $v_k$ in $S$ are also distinct.  Similarly, if $v_j \in S'$, the distances from $v_j$ to vertices between $v_1$ and $v_j$ in $S'$ would be distinct and so would those from $v_j$ to vertices between $v_j$ and $v_k$ in $S'$.  \\ 

\noindent Consider any vertex $v_i$ in $S$.  From the discussion in the previous paragraph, the number of vertices in $S$ that are of unit distance from $v_i$ is at most two.  Therefore, the number of unit isosceles triangles with apex vertex $v_i$ that are completely within $S$ is one, and hence there are at most $|S| = a$ unit isosceles triangles in $S$.  A similar result holds for $S'$.  Now, consider the number of unit isosceles triangle with its base completely within $S$.  There are $(a^2 - a)/2$ bases in $S$, and for each one, its perpendicular bisector can intersect $S'$ in at most one place (or else convexity would be contradicted).  Hence, there are at most $(a^2 - a)/2$ unit isosceles triangles with their bases completely in $S$.  A simlar result holds for $S'$.  \\ 

\noindent Finally, consider unit isosceles triangles such that a vertex of the base and the apex vertex are either both in $S$ or $S'$.  Suppose both are in $S$.  For any vertex $v\in \mathcal{P}$, define $g_{s}(v)$ to be the number of vertices in a subset $s\in \mathcal{P}$ that are of unit distance from $v$ and let $g_{\mathcal{P}} (v) = g(v)$.  Then, by the arguments given above, for any $v\in S$, $g_{S} (v) \le 2$.  Suppose that two such vertices exist, namely $v_1$ and $v_2$ such that $d(v_1,v )= d(v_2,v) = 1$.  Then, $\triangle v_1vu$ is isosceles if and only if $d(v, u) = 1$, so there are at most $2g(v)$ unit isosceles triangles with apex vertex $v$ such that the base is within $S$.  Summing over all $v\in \mathcal{P}$, we attain that the number of unit isosceles triangles of the above type is at most $2\sum_{v\in \mathcal{P}} g(v)$, which corresponds to four times the number of unit distances in a convex $n$-gon. In [2], Aggarwal proved that there are at most $n\log_2 n + 4n$ unit distances in a convex $n$-gon, and hence, there are at most $4n\log_2 n + 16n$ triangles of the this form.  \\

\noindent Upon summing, we attain that there are at most
\begin{center}
 $\displaystyle\frac{a^2 + b^2 - a - b}{2} + a + b + 4n\log_2 n + 16n  < \displaystyle\frac{(a+b+1)^2 }{2} + 4n\log_2 n + 16n $ \\ 
 $\le \displaystyle\frac{(n+3)^2}{2} + 4n\log_2 n + 16n < \displaystyle\frac{n^2}{2} + 4n\log_2 n + 19n + \displaystyle\frac{9}{2}$
 \end{center} 
 \noindent unit isosceles triangles for sufficiently large $n$. \\ 

\noindent {\bfseries Case 2:} There are at least three vertices of $\mathcal{P} = v_1v_2v_3\cdots v_n$ on the circle.  Three of these vertices must form an acute triangle, say $v_1, v_x, v_y$ with $1\le x\le y\le n$.  Let the vertices $v_1, v_2, \cdots v_x$ form $S_1$, the vertices $v_xv_{x+1}v_{x+2}\cdots v_y$ form $S_2$, and the vertices $v_{y}v_{y+1}v_{y+2}\cdots v_nv_1$ form $S_3$.  Let $|S_1| = a$, $|S_2| = b$, and $|S_3| = c$.  Since all vertices of $\mathcal{P}$ lie in the region defined by the union of $\triangle v_1v_xv_y$, and the semicircles with diameters $v_1v_x$, $v_xv_y$, and $v_1v_y$, $a + b + c \le n + 3$.  We proceed in a similar manner as before. Again count the total number of isosceles triangles included only in $S_1$, in only $S_2$, and only in $S_3$.  By using the same argument as applied in Case 1, this number is at most $a + b + c \le n + 3$.  Also, by using the same reasoning as provided in Case 1, the number of unit isosceles triangles with a vertex of the base and the apex vertex in the same set is at most $n\log_2 n + 4n$.  Now, we consider the case in which each vertex of the base is in a different set from the set in which the apex vertex resides.  Suppose both vertices of the base lie in $S_1$.  Then, there are $(a^2-a)/2$ possible bases and the perpendicular bisector can hit $\mathcal{P} - S_1$ in at most one place, thereby yielding at most $(a^2-a)/2$ possible unit isosceles triangles with the base exclusively in $S_1$.  Similar results hold for $S_2$ and $S_3$.  Next, the case in which a base has one vertex in $S_2$ and the other in $S_3$. There are at most $bc$ such bases, and since each of their perpendicular bisectors can only hit $S_1$ in one place, there are at most $bc$ unit isosceles triangles with a base partly in $S_2$ and partly in $S_3$.  Similar results hold for the others cases. \\

 \noindent Upon summing these four quantities, the number of unit isosceles triangles is at most
\begin{center}
 $\displaystyle\frac{a^2 + b^2 + c^2}{2} + ab + bc + ac + \displaystyle\frac{  a + b + c}{2} + 4n\log_2 n + 16n$ \\ 
 $< \displaystyle\frac{(a + b + c + 1)^2}{2} + 4n\log_2 n + 16n< \displaystyle\frac{(n + 4)^2}{2} + 4n\log_2 n + 16n< \displaystyle\frac{n^2}{2} + 4n\log_2 n + 20n + 8$ \\ 
 \end{center} \noindent for sufficiently large $n$.  \rule[0mm]{2.6mm}{2.6mm}% 
 	
\paragraph{Remark:} If the number of unit distances in a convex polygon can be shown to be at most $\Theta(n)$, then the number of unit isosceles triangles can be proven to be at most $n^2/2+ cn$ for a suitable constant $c$.  
 \section{Number of General Isosceles Triangles}
 \paragraph{Conjecture 1:} Let $I(n)$ denote the maximum possible number of isosceles triangles formed by verticees of a convex $n$-gon, with $n\ge 3$. Then, $I(n) \le \displaystyle\frac{3n^2}{4} + \Theta(n)$.    
 \subsection{Preliminary Observations}
 
 \paragraph {Proposition 2:} $I(n) \ge \displaystyle\frac{1}{4} (3n^2 - 11n + 8 + 2\left\lfloor \displaystyle\frac{n}{2} \right\rfloor)$. 

\paragraph {Proof:} Suppose $n$ is even (the logic is identical for odd $n$) and let $n = 2x$.  Consider vertices $v_1, v_2, v_3, \cdots , v_{n-1}$ on a circle with center $v$ such that the polygon $vv_{n-1}v_{n-2}\cdots v_1$ is convex and $d(v_i, v_{i+1}) = d(v_{j}, v_{j+1})$ for all $1\le i, j\le n - 2$.  Then, $\triangle vv_iv_j$ is an isosceles triangle for all $1\le i < j \le n$, and hence we obtain $(n^2 - 3n + 2)/2$ such isosceles triangles.  Moreover, $\triangle v_{i - 1}v_i v_{i+1}, \triangle v_{i - 2}v_iv_{i + 2}, \cdots , \triangle v_{1}v_iv_{2i - 1}$ are all isosceles for any integer $1\le i\le x$.  Hence, $v_i$ is the apex vertex of $i - 1$ isosceles triangles.  Summing this over $1\le i \le x$ gives $\sum_{i = 1}^{x} (i - 1) = (x^2 - x)/2$ isosceles triangles with apex vertex being among the set $\{ v_1, v_2, v_3, \cdots , v_{k} \}$.  Moreover, $v_{n - i}$ is the apex vertex of $i - 1$ isosceles triangles for $n - x - 1\le i \le n$, forming another $\sum_{i=1}^{n - x - 1} (i - 1) = (n-x-1)(n-x-2)/2$ isosceles triangles.  Summing all three quantities yields the number of isosceles triangles to be $(3n^2 - 10n + 8)/4$.  When $n$ is odd, we can repeat the same process and attain $(3n^2 - 10 + 7)/4$ isosceles triangles. \rule[0mm]{2.6mm}{2.6mm}% 

\paragraph{Definitions:} Define the {\itshape apex vertex} of an isosceles triangle to be the vertex common to both legs of the triangle and say that an isosceles triangle {\itshape belongs} to its apex vertex.  Let $T(\mathcal{P})$ denote the number of isosceles triangles in a convex polygon $\mathcal{P}$, and note that $I(n) = \max(T(\mathcal{P}))$ over all convex $n$-gons $\mathcal{P}$. 

\paragraph {Proposition 3:} Suppose $\mathcal{P}$ does not have any centroids.  Then $T(\mathcal{P}) \le n \left\lfloor \displaystyle\frac{n - 1}{2} \right\rfloor $.  

\paragraph {Proof:} We in fact show that no vertex can be the apex vertex of more than $\lfloor (n-1)/2 \rfloor$ isosceles triangles, which proves the desired bound.  Suppose that some vertex $v$ is the apex vertex of more than $(n-1)/2$ isosceles triangles.  There are two base vertices for each isosceles triangle, which entails that there are more than $n - 1$ base vertices total.  However, there are $n - 1$ vertices other than $v$ in $\mathcal {P}$, and hence one vertex $u$ is used in two isosceles triangles.  Suppose that $\triangle vut$ and $\triangle vus$ are isosceles.  Then, $d(v,s) = d(v,u) = d(v,t)$, so $v$ is a centroid, which is impossible, thereby proving proposition 3. Note that equality holds when $\mathcal{P}$ is a regular polygon. \rule[0mm]{2.6mm}{2.6mm}% 

\subsection{Proof for Non-Intersecting Centroid-Circles}
 
 \paragraph{Theorem 2:} Let $\mathcal {P}$ be a convex $n$-gon composed of $k$ non-intersecting centroid-circles; then, $T(\mathcal {P})\le 3(n+1)^2/4$.  

\paragraph{Proof:} We prove $T(\mathcal{P}) \le 3n^2/4$ when $n$ is even, which implies that $T(\mathcal{P}) \le 3(n+1)^2/4$ when $n$ is odd.   Suppose that the centroid-circles are $C_1, C_2, C_3, \cdots , C_k$,  suppose that $C_i$ has $a_i$ vertices on its circle, and without loss of generality, suppose that $a_1\ge a_2\ge a_3\ge \cdots \ge a_k$.  Since the centroids do not intersect, $\sum_{i = 1}^k a_i \le n$.  Let $v_{ij}$ be the $j$th vertex in counterclockwise order on the circle of $C_i$.  Consider two cases: one in which there is a centroid-circle with more than $n/2$ vertices and the other in which there is not. \\ \\ {\bfseries Case 1:} $a_1 > \displaystyle\frac{n}{2}$, so $a_1 > \displaystyle\sum_{i = 2}^k a_i$.  \\ \\ Set $a_1 - n/2 = x$.  Let $S_1$ consist of $v_{1i}$ for $1\le i \le x$.  Let $S_2$ consist of $v_{1i}$ for $x + 1\le j \le n - x$, and let $S_3$ consist of $v_{1i}$ for $n - x + 1 \le i \le n$.  Note that the number of isosceles triangles due to all centroids is at most $\sum_{i=1}^k (a_i^2 - a_i)/2$.  Take some $v_{1j}$ in $S_1$ that is not a centroid.  $v_{1j}$ cannot be on the perpendicular bisector of the segment formed by two vertices on the circle of $C_1$ between $v_{11}$ and $v_{1(j -1)}$, or else since $C_1$ also lies on this perpendicular bisector, contradicting convexity.  Moreover, notice that $v_{1i}$ cannot be part of two triangles with apex vertex $v_{1j}$, or else $v_{1j}$ is a centroid by the logic used in proposition 3.  Hence, $v_{1j}$ can have at most $j - 1$ isosceles triangles having a vertex on $C_1$ between $v_{11}$ and $v_{1(j - 1)}$.   Through similar reasoning, $v_{1j}$ cannot be the apex vertex of a triangle with the two base vertices in $S_3$, so at least one of the vertices in any isosceles triangle with apex vertex $v_{1j}$ that does not have a vertex between $v_{11}$ and $v_{1j}$ has a vertex in $C_2, C_3, \cdots , C_k$, which has cardinality $n - a_1$.  Again, no vertex among these can be in two isosceles triangles with apex vertex $v_{1j}$, implying that $v_{1j}$ is an apex vertex of at most $n - a_1$ triangles having a base not entirely within $S_1$; as a result, $v_{1j}$ is an apex vertex of at most $n - a_1 + j - 1$ triangles.  Analagously, if we take $v_{1(n - j)}$, for $j\le x - 1$, at most $n - a_1 + j$ isosceles triangles can be formed.  Summing this over all vertices in $S_1$ and $S_3$ yields at most $2\sum_{i = 1}^x (n - a_1 + j) \le 2x(n - a_1) + x^2$ isosceles triangles.  Now, each of the vertices in $S_2$ or $\mathcal{P} - S_1 - S_2 - S_3$ can be the apex vertex of at most $n/2$ isosceles triangles, totalling $n(n - a_1)$ isosceles triangles.  Suppose that $n - a_1 = b$.  Summing the four quantities yields a total of 
\begin{eqnarray*}
(a_1 + b)b + \frac{(a_1 - b)^2}{4} + (a_1 - b)b + \displaystyle\sum_{i = 1}^k \frac{a_i (a_i - 1)}{2} < \frac{a_1^2 + 6a_1b + b^2}{4} + \frac{a_1^2 + b^2}{2} = \frac{3}{4} \cdot n^2
  \end{eqnarray*} 
  isosceles triangles. \\ \\ \\ \noindent {\bfseries Case 2}: $a_i\le \displaystyle\frac{n}{2}$ for all $1\le i\le k$. \\ \\ The centroid vertices give at most $\sum_{i = 1}^k (a_i^2-a)/2$ isosceles triangles.  Any non-centroid vertex of $\mathcal {P}$ can form at most $\lfloor (n-1)/2 \rfloor$ isosceles triangles by proposition 3, so the total number of isosceles triangles formed by non-centroid vertices is at most $n\lfloor (n-1)/2 \rfloor $, and hence the total number of isosceles triangles is at most 
  \begin{center}
  $n\left\lfloor \displaystyle\frac{n-1}{2} \right\rfloor + \displaystyle\sum_{i = 1}^k \displaystyle\frac{a_i^2-a_i}{2} < \displaystyle\frac{n^2}{2} + \displaystyle\sum_{i=1}^k \displaystyle\frac{a_i^2}{2}$
  \end{center}  Since the function $f(x) = x^2$ is convex and $a_i \le n/2$ for $1\le i\le k$, $\sum_{i=1}^k a_i^2$ is maximized when $a_1 = a_2 = n/2$, yielding the number of isosceles triangles to be less than $3n^2/4$. \rule[0mm]{2.6mm}{2.6mm}%  \\ 

\subsection{Potential Progress Towards Intersecting Centroid-Circles} 
We omit the proof of the following partial result:
\paragraph{Theorem 3:} Suppose that $\mathcal{P}$ is a convex $n$-gon that has $k$ intersecting centroid-circles with $k = o(n^{2/3})$. Then, there are at most $3n^2/4 + o(n^2)$ isosceles triangles formed by vertices of $\mathcal{P}$. 
\subsection{Number of Distinct Distances From a Vertex}
\paragraph{Proposition 4:} In a convex polygon $\mathcal{P} = v_1v_2v_3\cdots v_n$, let $d(v_i)$ be the number of distinct lengths among $v_1v_i, v_2v_i, \cdots , v_{n}v_i$.  Let $d(\mathcal{P}) = \displaystyle\max_{1\le i\le n} d(v_i)$.  If $I(n) \le 3n^2/4 + o(n^2)$, then $d(\mathcal{P}) \ge 5n/12 + o(n)$.  

\noindent \paragraph{Proof:} The method is identical to that of Dumitrescu given in [7]. Let $I(v)$ be the number of triangles a vertex $v\in \mathcal{P}$ is an apex of.  Then, $\sum_{v\in \mathcal{P}} I(v) = T(\mathcal{P}) \le 3n^2/4 + o(n^2)$. Let $k$ be the maximum number of distinct lengths coming from a single vertex.  As noted by Dumitrescu, $T(\mathcal{P})$ is minimized when, for each vertex $v\in \mathcal{P}$, the other $n - 1$ vertices distributed evenly on concentric circles centered at $v$, i.e., each circle contains either $2$ or $3$ vertices.  Let there be $x$ circles with $2$ vertices and $y$ circles with $3$ vertices about some vertex $v$.  Then, $2x + 3y = n - 1$ and $x + y \le k$.  Therefore, $x \le 3k - n + 1$, thus $I(v) = x + 3y \ge 2n - 2 - 3k$.  Consequently, $3n^2/4 + o(n^2) \ge I(\mathcal{P}) \ge n(2n - 2 - 3k)$, and so $k\ge 5n/12 + o(n)$. \rule[0mm]{2.6mm}{2.6mm}% 
 \section{Number of Regular Polygons}
 Pach and Pinchasi proved in [16] that there are at most $\lfloor 2(n-1)/3 \rfloor $ unit equilateral triangles in a convex $n$-gon, whereas in [1], \'{A}brego and Fern\'{a}ndez-Merchant provided an upper bound of $n - 2$ (not necessarily unit) equilateral triangles.  However, the precise bound on the number of  equilateral triangles remains open.  Therefore, we believe   
 \paragraph{Conjecture 2:} The maximum number of equilateral triangles in a convex $n$-gon is at most $\lfloor 2(n-1)/3 \rfloor$.  
 
 \paragraph{Remark:} Notice that $\lfloor 2(n-1)/3 \rfloor$ equilateral triangles are formed in the following position: let vertices $v, v_1, v_2, v_3, \cdots v_{n-1}$ be such that $v_1v_2\cdots v_n v$ is convex, $d(v, v_i) = 1$ for all $1\le i \le n -1$, and $d(v_i, v_{i+k}) = d(v_{i+k}, v_{i+2k}) = 1$, where $k = \lfloor n/3 \rfloor$ and $1\le i\le \lfloor (n-1)/3 \rfloor$. Then, $\triangle vv_{i+k}v_{i+2k}$ and $\triangle vv_iv_{i+k}$ are equilateral for all $1\le i\le \lfloor (n-1)/3 \rfloor$, which gives a total of $\lfloor 2(n-1)/3 \rfloor$ equilateral triangles. This configuration has been mentioned by both \'{A}brego and Fern\'{a}ndez-Merchant in [1] and Pach and Pinchasi in [16]. \\
 
 \noindent While we are unable to prove conjecture 2, we are able to find precise bounds for the number of regular $k$-gons in a convex $n$-gon, for $k\ge 4$. 
 
 \paragraph{Theorem 4:} Let $n$ and $k$ be integers greater than $3$.  The maximum number of regular $k$-gons that can be found in a convex $n$-gon is $\lfloor n/k \rfloor$ and this bound is sharp.
 
 \paragraph{Proof:} We first show that equality can be achieved.  Let $n = qk + r$, where $0 \le r < k$.  Consider a regular $qk$-gon $v_1v_2v_3\cdots v_{qk}$ and place the other $r$ vertices on the circumcircle of the $k$-gon arbitrarily.  The polygon formed is convex and for any $1\le i \le q$, $v_iv_{q+i}v_{2q + i}\cdots v_{kq - q + i}$ is a regular $k$-gon and hence there are $q = \lfloor n/k \rfloor$ regular $k$-gons in this polygon. \\ 
 
 \noindent  We now  prove the upper bound.  Let the polygon be $v_1v_2v_3\cdots v_n$.  Let the {\itshape degree} of a vertex denote the number of regular $k$-gons that pass through that vertex and call two polygons {\itshape disjoint } if the intersection between the two polygons has area zero.  We show that the degree of any vertex is at most one.  Suppose to the contrary that a vertex, say $v_1$, has degree at least two.  Consider the largest regular $k$-gon passing through $v_1$.  Let it be $\mathcal{U} = u_1u_2\cdots u_{k-1}v_1$.  Let one of the other regular $k$-gons be $t_1t_2t_3\cdots t_{k-1}v_1$.  If these polygons are disjoint, then $ \angle t_1v_1t_{k-1} = \pi - (2\pi/k) = \angle u_1v_1u_{k-1}$, and thus there is an angle of at least $2\pi -(4\pi/k) \ge \pi$ since $k \ge 4$, which contradicts convexity.  Consequently, these $k$-gons are not disjoint and so, for some $i$, $v_1t_i$ passes through $\mathcal{U}$.  Let $v_1t_i$ hit $\mathcal{U}$ again at $t$ and suppose $t$ lies on $u_ju_{j+1}$ for some $j$.  Then, $d(v_1, t) < d(v_1, t_i) \le d(v_1, u_1) = d(v_1, u_{k-1})$.  \\ 
 
 \noindent Now, if $\angle v_1u_1u_j \ge \pi/2$ and $\angle v_1u_1u_j \ge \pi/2$ (or they are both at most than $\pi/2$, in which case rather than considering $u_1$, consider $u_{k-1}$), then $\angle v_1u_1t \ge \pi/2$, so $v_1t > v_1u_1$, which is a contradiction.  Otherwise, $j$ is the unique vertex such that $\angle v_1u_1u_j \le \pi/2$ and $\angle v_1u_1u_j \ge \pi/2$, hence $j = \lfloor k/2 \rfloor$.  In this case, $\angle v_1u_1t \ge \angle v_1u_1u_{j+1}$ and $\angle u_1v_1t \ge \angle u_1v_1u_j$, thus $\angle v_1u_1t + \angle u_1v_1u_j \ge \pi - (2\pi/k)$, so $\angle u_1tv_1 \le 2\pi/k$, implying that if $k - j - 1\ge 2$, $\angle tu_1v_1 \ge 2\pi/k \ge \angle u_1tv_1$.  This implies that $d(v_1, u_1) \le d(v_1, t)$, which is a contradiction.  Consequently, $k = 4$, so $\angle v_1u_1t = \pi/2$, entailing that $d(v_1, t) > d(v_1, u_1)$, which is a contradiction. \\
 
 \noindent Therefore, every vertex has degree at most one, and as a result, the sum of the degrees is at most $n$.  However, every $k$-gon has $k$ vertices, each having degree one, so there are at most $\lfloor n/k \rfloor$ regular $k$-gons.   \rule[0mm]{2.6mm}{2.6mm}% 

\section{Polygons With Unit Perimeters}
In this section, we use theorems of Altman given in [4] to prove Audet, Hansen, and Messine's conjecture given in [5]. In 2008, in [13], Larger and Pillichshammer also prove this conjecture. Here, we give a simpler proof. 
\paragraph{Theorem 5:} For any convex $n$-gon with unit perimeter, the sum $S_n$ of distances between its vertices satisfies \begin{center} $\displaystyle\frac{n-1}{2} \le S_n \le \displaystyle\frac{1}{2}\cdot  \left\lceil \displaystyle\frac{n}{2} \right\rceil \left\lfloor \displaystyle\frac{n}{2} \right\rfloor$ \end{center}
\paragraph{Proof:} Let the polygon be $v_1v_2v_3\cdots v_n$ and let $\displaystyle\sum_{i=1}^n d(v_i, v_{i+j}) = u_j$ (where indices are taken modulo $n$).  In his first theorem in [4], Altman shows that $u_i < u_j$ whenever $1\le i < j \le \lfloor n/2 \rfloor$.  Since $u_1$ is the perimeter of the polygon, $u_j \ge u_1 = 1$ for all $1\le j\le \lfloor n/2 \rfloor$.  Moreover, notice that, for any $i$ and any $j, k \le \lfloor n/2 \rfloor$, $d(v_i, v_{i+j}) + d(v_{i+j}, v_{i+k+j}) > d(v_{i}, v_{i+j+k})$ by the triangle inequality.  Summing over all $i$ yields $u_j + u_k > u_{j+k}$.  In particular, $u_2 < 2u_1 = 2$, and by induction, $u_i < i$ for all $1\le i\le \lfloor n/2 \rfloor$.  Observe that $S_n = \displaystyle\sum_{i=1}^{(n-1)/2} u_i$ when $n$ is odd and $S_n = \displaystyle\sum_{i=1}^{(n-2)/2} u_i + (u_{n/2})/2$ when $n$ is even.  \\
\noindent Therefore, if $n$ is odd, then the following two inequalities hold:
\begin{center} $S_n = \displaystyle\sum_{i=1}^{\frac{n-1}{2}} u_i \ge \sum_{i=1}^{\frac{n-1}{2}} 1 = \displaystyle\frac{n-1}{2}; \quad S_n = \displaystyle\sum_{i=1}^{\frac{n-1}{2}} u_i \le \displaystyle\sum_{i=1}^{\frac{n-1}{2}} i = \displaystyle\frac{(n-1)(n+1)}{8} = \displaystyle\frac{1}{2}\cdot \left\lfloor \frac{n}{2} \right\rfloor \left\lceil \frac{n}{2} \right\rceil$
\end{center} thereby, proving the theorem.  Analogously, if $n$ is even, 
\begin{center} $S_n = \displaystyle\sum_{i=1}^{\frac{n}{2} - 1} u_i + \displaystyle\frac{u_{n/2}}{2} \ge \displaystyle\sum_{i=1}^{\frac{n}{2} - 1} 1 + \displaystyle\frac{1}{2} = \displaystyle\frac{n-1}{2}; \quad S_n = \displaystyle\sum_{i=1}^{\frac{n-2}{2}} u_i + \displaystyle\frac{u_{n/2}}{2} \le \displaystyle\sum_{i=1}^{\frac{n-2}{2}} i + \displaystyle\frac{n}{4} = \displaystyle\frac{n^2}{8}$ 
\end{center} thereby, proving the theorem. \rule[0mm]{2.6mm}{2.6mm}% 
\paragraph{Remark:} Audet, Hansen, and Messine have already shown that the lower bound is approached with a segment $[0, 1/2]$ with $v_1$ at $0$, and $v_2 , v_3 , ..., v_n$ arbitrarily close to $1/2$ and the upper bound is approached with $v_1 , v_2, ...., v_{\lfloor n/2 \rfloor}$ arbritraliy close to $0$, and $v_{\lfloor n/2+1 \rfloor} , ..., v_n$ arbitrarily close to $1/2$.

\end{document}